\documentclass[twocolumn,prl,showpacs,amsmath,assymb,eufrak]{revtex4}
\usepackage{epsfig,amsfonts,amsbsy}

\makeatletter
\long\def\@makecaption#1#2{{\small
\advance\leftskip1cm
\advance\rightskip1cm
\vskip\abovecaptionskip
\sbox\@tempboxa{#1: #2}%
\ifdim \wd\@tempboxa >\hsize
 #1: #2\par
\else
\global \@minipagefalse
\hb@xt@\hsize{\hfil\box\@tempboxa\hfil}%
\fi
\vskip\belowcaptionskip}}
\makeatother
\def\eq#1\en{\begin{equation}#1\end{equation}}  
\def\eqa#1\ena{\begin{align}#1\end{align}}
\def\eqg#1\eng{\begin{gather}#1\end{gather}}
\newcommand{\lb}[1]{\label{e:#1}}
\newcommand{\rlb}[1]{\eqref{e:#1}} 
\newcommand{\nl}{\notag\\}

\newcommand{\norms}[1]{\Vert#1\Vert}

\newcommand{\sbkt}[1]{\langle#1\rangle}

\newcommand{\sumtwo}[2]%
{\mathop{\sum_{#1}}_{#2}}
\newcommand{\sumthree}[3]%
{\mathop{\mathop{\sum_{#1}}_{#2}}_{#3}}
\newcommand{\sumfour}[4]%
{\mathop{\mathop{\mathop{\sum_{#1}}_{#2}}_{#3}}_{#4}} 
\newcommand{\prodtwo}[2]%
{\mathop{\prod_{#1}}_{#2}}
\newcommand{\mintwo}[2]%
{\mathop{\min_{#1}}_{#2}}
\newcommand{\maxtwo}[2]%
{\mathop{\max_{#1}}_{#2}}
\newcommand{\maxthree}[3]%
{\mathop{\mathop{\max_{#1}}_{#2}}_{#3}}
\newcommand{\limtwo}[2]%
{\mathop{\lim_{#1}}_{#2}}
\newcommand{\suptwo}[2]%
{\mathop{\sup_{#1}}_{#2}}
\newcommand{\supthree}[3]%
{\mathop{\mathop{\sup_{#1}}_{#2}}_{#3}}
\newcommand{\supfour}[4]%
{\mathop{\mathop{\mathop{\sup_{#1}}_{#2}}_{#3}}_{#4}} 
\newcommand{\inftwo}[2]%
{\mathop{\inf_{#1}}_{#2}}
\newcommand{\infthree}[3]%
{\mathop{\mathop{\inf_{#1}}_{#2}}_{#3}}
\newcommand{\inffour}[4]%
{\mathop{\mathop{\mathop{\inf_{#1}}_{#2}}_{#3}}_{#4}} 

\newcommand\calH{{\cal H}}
\newcommand\calI{{\cal I}}

\newcommand{\mrx}{\mathrm{x}}
\newcommand{\mry}{\mathrm{y}}
\newcommand{\mrz}{\mathrm{z}}

\newcommand{\bsn}{\boldsymbol{n}}

\newcommand{\bsr}{\boldsymbol{r}}

\newcommand{\bbR}{\mathbb{R}}

\newcommand{\La}{\Lambda}
\newcommand{\pmz}{\{+,-,0\}}
\newcommand{\ad}{a^\dagger}
\newcommand{\Stot}{S_\mathrm{tot}}
\newcommand{\PGS}{\Psi^\mathrm{GS}}
\newcommand{\EGS}{E^\mathrm{GS}}
\newcommand{\Phin}{\Phi_{\bsn}}
\newcommand{\midskip}{\vspace{5pt}}

\begin{document}
\title{Ground States of the Spin-1 Bose-Hubbard Model}

\author{Hosho Katsura and Hal Tasaki}
\affiliation{Department of Physics, Gakushuin University, Mejiro, Toshima-ku, Tokyo 171-8588, Japan}

\date{\today}

\begin{abstract}
We prove basic theorems about the ground states of the $S=1$ Bose-Hubbard model.
The results are quite universal and depend only on the coefficient $U_2$ of the spin-dependent interaction.
We show that the ground state exhibits saturated ferromagnetism if $U_2<0$, is spin-singlet if $U_2>0$, and exhibits ``SU(3)-ferromagnetism'' if $U_2=0$, and completely determine the degeneracy in each region.
\end{abstract}


\maketitle

Recent progress in cold atom experiments, in particular those of bosons trapped in an optical lattice, has led to a renewed interest in the low energy properties of the Bose-Hubbard model \cite{BH1-1, BH1-2, BH_coldatom}.
Especially the system of spinor bosons, in which hyperfine spin degrees of freedom couple to many-body quantum physics of bosons, are expected to have a variety of phases including the spin-singlet, nematic, and ferromagnetic phases \cite{Yip, Imambekov, Snoek,Turner, Semenoff}.
From a theoretical point of view, it is also interesting to compare the situation with that of the Fermi-Hubbard model, where intricate interplay between the spin degrees of freedom and many-body quantum physics has been studied in depth \cite{Hubb_review_Lieb,Hubb_review_Hal}.

Here we shall present basic theorems about the ground states of the standard Bose-Hubbard model for $S=1$ bosons.
We precisely characterize, in each parameter range, the degeneracy and the total spin angular momentum of the ground states, and identify the signs of the ``wave function'' represented in an appropriate basis.
The result depends only on the single parameter $U_2$, and one gets ferromagnetic ground states for $U_2<0$, a spin-singlet ground state  for $U_2>0$ (when the boson number is even), and ground states with ``SU(3)-ferromagnetism'' for $U_2=0$.
These results are surprisingly universal, and do not depend on the lattice structure, the spin-independent interaction, or the boson number.
This is in a sharp contrast with corresponding results in the Fermi-Hubbard model \cite{Hubb_review_Lieb,Hubb_review_Hal,LiebsTHeorem} and the Heisenberg spin system \cite{Marshall,LM}, where strict restrictions on lattice structures and the electron numbers are usually required.

We hope that these basic and rigorous results will be useful for the future theoretical and numerical studies of the Bose-Hubbard model.
We note that our Theorem~3 for $U_2=0$ is a special case of the theorem by Eisenberg and Lieb proved for spinor bosons in continuum \cite{EisenbergLieb}.

\midskip
\paragraph*{Definitions and main theorems:}
We consider a system of $N$ spinor bosons with $S=1$ on a finite lattice $\La$, where $N$ is arbitrary and fixed.
For each site $x\in\La$, we denote by $\ad_{x,\sigma}$ and $a_{x,\sigma}$ the creation and the annihilation operators, respectively, of a boson at $x$ with spin $\sigma\in\pmz$.
We define the number operators by $n_{x,\sigma}:=\ad_{x,\sigma}a_{x,\sigma}$ and $n_x:=\sum_{\sigma\in\pmz}n_{x,\sigma}$, and the spin operators by $S^{(\alpha)}_x:=\sum_{\sigma,\tau\in\pmz}\ad_{x,\sigma}\,S^{(\alpha)}_{\sigma,\tau}\,a_{x,\tau}$ for $\alpha=\mrx,\mry,\mrz$, where $S^{(\alpha)}_{\sigma,\tau}$ are the spin matrices for $S=1$.
We write $\mathbf{S}_x:=(S^{(\mrx)}_x,S^{(\mry)}_x,S^{(\mrz)}_x)$.
The total spin operators are $\Stot^{(\alpha)}:=\sum_{x\in\La}S^{(\alpha)}_x$.
As usual we define $\Stot^\pm=\Stot^{(\mrx)}\pm i\Stot^{(\mry)}$, and write the eigenvalue of $(\mathbf{S}_\mathrm{tot})^2=\sum_{\alpha=\mrx,\mry,\mrz}(\Stot^{(\alpha)})^2$ as $\Stot(\Stot+1)$.

We shall make a frequent use of the basis states
\eq
\Phin:=\bigl\{
\hspace{-5pt}
\prod_{x\in\La,\,\sigma\in\pmz}
\hspace{-5pt}
(\ad_{x,\sigma})^{n_{x,\sigma}}
\bigr\}\,\Phi_\mathrm{vac},
\lb{Phin}
\en
where $\Phi_\mathrm{vac}$ is the state with no bosons in the trap, and the multi-index $\bsn=(n_{x,\sigma})_{x\in\La,\,\sigma\in\pmz}$ is a collection of nonnegative integers such that $\sum_{x,\sigma}n_{x,\sigma}=N$.
The whole Hilbert space $\calH$ is spanned by all such $\Phin$.
For $M=0,\pm1,\ldots,\pm N$, we denote by $\calI_M$ the set of $\bsn$ satisfying $\sum_x(n_{x,+}-n_{x,-})=M$, and by $\calH_M$ the subspace of $\calH$ spanned by $\Phin$ with $\bsn\in\calI_M$.
Note that any $\Psi\in\calH_M$ satisfies $\Stot^{(\mrz)}\Psi=M\Psi$, and that $\calH=\bigoplus_{M=-N}^N\calH_M$.

We consider the standard Hamiltonian of the $S=1$ Bose-Hubbard model \cite{Imambekov, Snoek}
\eqa
H=&-
\hspace{-10pt}
\sum_{x,y\in\La,\,\sigma\in\pmz}
\hspace{-10pt}
t_{x,y}\,\ad_{x,\sigma}a_{y,\sigma}
+\sum_{x\in\La}V_x\,n_x
\nl&+\sum_{x\in\La}\biggl[
U_0\frac{n_x(n_x-1)}{2}+
U_2\Bigl\{\frac{(\mathbf{S}_x)^2}{2}-n_x\Bigr\}
\biggr],
\lb{H}
\ena
which has a global SU(2) symmetry of the spin rotation.
We assume that the hopping matrix elements $t_{x,y}=t_{y,x}$ is nonnegative, and the whole lattice $\La$ is connected via nonvanishing $t_{x,y}$.
The standard choice, where one sets $t_{x,y}=t>0$ for neighboring $x$, $y$, and $t_{x,y}=0$ otherwise, is sufficient.
Here $V_x\in\bbR$ is the on-site single-particle potential, and $U_0$ and $U_2$ are real coefficients for the spin-independent and spin-dependent two-body interactions, respectively.
We make no assumptions on $V_x$ and $U_0$.
Assumptions on $U_2$ are stated in the theorems.

In an (idealized) experimental realization of the Bose-Hubbard model, one  initially prepares specified numbers of bosons with $+$, 0, and $-$ spins in the trap, and lets the system evolve in time.
As long as $U_2\ne0$, spin flips take place and the state evolves into a superposition of various spin states,
but the eigenvalue of $\Stot^{(\mrz)}$, which is $M$, is conserved. 
Therefore the system is expected to evolve into a low energy state within each subspace $\calH_M$.
It thus makes sense to consider local ground states within each  $\calH_M$.

{\em Theorem 1}\/---
If $U_2<0$, the local ground state $\PGS_M$ in $\calH_M$ is unique, and is written as
\eq
\PGS_M=\sum\nolimits_{\bsn\in\calI_M}\alpha_{\bsn}\Phin,
\lb{GS1}
\en
with $\alpha_{\bsn}>0$, and has the maximum possible total spin $\Stot=N$.
The local ground state energy $\EGS_M$ is independent of $M$.
Thus each $\PGS_M$ is the global ground state in $\calH$, and the ground states are $(2N+1)$-fold degenerate.

In short the ground states exhibit saturated ferromagnetism, and are unique apart from the trivial $(2\Stot+1)$-fold degeneracy.
Note that the ground states of the present bosonic system with $U_2<0$ may be in various phases, including Bose-Einstein condensation and Mott insulator.
As for the magnetic property, however, the ground states universally exhibit saturated ferromagnetism.

{\em Theorem 2}\/---
If $U_2>0$, the local ground state $\PGS_M$ in $\calH_M$ is unique, and is written as
\eq
\PGS_M=\sum\nolimits_{\bsn\in\calI_M}\beta_{\bsn}\,(-1)^{\sum_{x}n_{x,0}/2}\,\Phin,
\lb{GS2}
\en
with $\beta_{\bsn}>0$.
The state $\PGS_M$ has total spin $\Stot=|M|$ if $N-M$ is even, and $\Stot=|M|+1$ if $N-M$ is odd.
The local ground state energy satisfies $\EGS_M=\EGS_{-M}$ for any $M$, $\EGS_{|M|}<\EGS_{|M|+1}$ if $N-M$ is even, and $\EGS_{|M|}=\EGS_{|M|+1}$ if $N-M$ is odd.

Thus $\PGS_0$ is the the unique global ground state in $\calH$ if $N$ is even, while $\PGS_1$, $\PGS_0$, and $\PGS_{-1}$ form a triplet of global ground states if $N$ is odd.
In short the ground states have the smallest total spin as possible, as is expected from the antiferromagnetic nature of the interaction.
We remark, however, that the present theorem is not sufficient to determine the magnetic property of the ground states, since the constraint on $\Stot$ alone does not specify magnetic structure uniquely.
Let us recall that the Heisenberg model on a bipartite lattice satisfies a theorem very similar to Theorem 2 \cite{Marshall,LM}, but it may exhibit various magnetic properties (e.g., antiferromagnetic long range order, quantum criticality, Haldane phase, dimerization) depending on details of the model.

Finally we focus on the special case $U_2=0$, where the spin-dependent interaction is absent. 
In this case no spin-flip takes place during the time evolution, and the numbers of $+$, 0, and $-$ spin bosons are conserved separately.
We thus have to talk about local ground states in smaller subspaces.
For three nonnegative integers $U$, $D$, and $Z$ such that $U+Z+D=N$, let us denote by $\calI_{U,D,Z}$ the set of $\bsn$ such that $\sum_xn_{x,+}=U$, $\sum_xn_{x,-}=D$, and $\sum_xn_{x,0}=Z$.
We also denote by $\calH_{U,D,Z}$ the subspace spanned by $\Phin$ with $\bsn\in\calI_{U,D,Z}$.

{\em Theorem 3}\/---
If $U_2=0$, the local ground state $\PGS_{U,D,Z}$ in $\calH_{U,D,Z}$ is unique, and is written as
\eq
\PGS_{U,D,Z}=\sum\nolimits_{\bsn\in\calI_{U,D,Z}}\gamma_{\bsn}\Phin,
\lb{GS3}
\en
with $\gamma_{\bsn}>0$.
The local ground state energy $\EGS_{U,D,Z}$ is independent of $({U,D,Z})$.
Thus each $\PGS_{U,D,Z}$ is the global ground state in $\calH$, and the ground states are $(N+1)(N+2)/2$-fold degenerate.

This theorem is a special lattice version of that proved in \cite{EisenbergLieb}, where details about the degeneracy is omitted.
See also \cite{MOR}.
The high degeneracy of the ground states can be understood  as a manifestation of the SU(3) symmetry of the model with $U_2=0$.
The ground state does not exhibit magnetic orderings in the standard sense, but there certainly is an ``exchange interaction'' which realizes the property \rlb{GS3}.
One may say that the ground states exhibit ``SU(3)-ferromagnetism.''

\midskip
\paragraph*{Limiting cases:}
While the above theorems provide only general information about the ground states, much more detailed information may be obtained in some special cases.

As a simple example, we show that one can precisely characterize the ground state in the ``singlet phase'' by using a rigorous perturbation theory.
Let $\La$ be a uniform lattice, such as the cubic lattice with periodic boundary conditions, and denote the number of sites as  $|\La|$.
We consider the case when $\nu:=N/|\La|$ is an even integer.
We set  $V_x=0$ for all $x\in\La$, and $t_{x,y}=t>0$ when $x$ and $y$ are neighboring sites and $t_{x,y}=0$ otherwise.
Then we shall perturb around the trivial model with $U_0>0$, $U_2>0$, and $t=0$.
By applying the general results (see Theorems~4.1 and 4.3) in \cite{Tom}, we can show the following.

{\em Theorem 4}\/---There are positive constants $B$ and $\delta$ which depend on the nature of $\La$ but not on the size $|\La|$.
When one has $U_0/t>B$ and  $U_2/t>B$, the unique ground state is a small perturbation of the product state where each site is occupied by $\nu$ bosons which form a spin-spin singlet.
There is an energy gap larger than $\delta$ above the ground state energy.

Let us also make a remark on the weak coupling limit, which essentially is a rewording of the earlier results in \cite{weak_Law,weak_Koahi_Ueda}.
Let us start from a model that satisfies the conditions for Theorems 1 or 2, and then take the limit $U_0,U_2\to0$.
Since the limiting theory is certainly that of free bosons, the (local) ground state in $\calH_0$ is written as
\eq
\Phi=\sum_{n}\frac{\eta_n}{\sqrt{n!}(\frac{N-n}{2})!}
(b^\dagger_0)^n(b^\dagger_+b^\dagger_-)^{(N-n)/2}
\Phi_\mathrm{vac},
\en
where we have assumed that $N$ is even.
Here $n$ is summed over even integers from $0$ to $N$, and $b^\dagger_\sigma$ is the creation operator of the unique single-particle ground state with spin $\sigma$.
Interestingly, the fact that  $\Phi$ has $\Stot=0$ or $N$ alone determines the coefficients $\eta_n$ uniquely.

We have (apart from normalization) $\eta_n=2^{n/2}/\{\sqrt{n!}\{(N-n)/2\}!\}$ if $\Stot=N$, and $\eta_n=(-2)^{-n/2}\sqrt{n!}/(n/2)!$ if $\Stot=0$ \cite{eta}.
Note that $|\eta_n|^2$ is proportional to the probability of observing $n$ spin-0 bosons in the state $\Phi$.
If $\Stot=N$, this probability distribution is $|\eta_n|^2\simeq\mathrm{const.}\exp[-2N\{(n/N)-(1/2)\}^2]$ for large $N$.
If $\Stot=0$, on the other hand, it becomes the power law distribution $|\eta_n|^2\simeq\mathrm{const.}/\sqrt{n}$, which is quite characteristic.

Of course exactly the same claim applies to a continuous model of bosons, i.e., bosons in  a standard optical trap (see the final part of the paper).
To our knowledge the power law distribution of the number of spin-0 bosons in the ground state with $\Stot=0$ has not yet been observed experimentally.

\midskip
\paragraph*{Proof of Theorem 1:}
We shall show below that $\sbkt{\Phin,H\Phi_{\bsn'}}\le0$ for any $\bsn$, $\bsn'$, and that the whole set $\calI_M$ are connected via nonvanishing $\sbkt{\Phin,H\Phi_{\bsn'}}$.
Then the Perron-Frobenius theorem \cite{PF} guarantees that the ground state in $\calH_M$ is unique and is written as \rlb{GS1}.

To determine the total spin and the degeneracy, note first that $\PGS_N$ (which consists only of spin $+$ bosons) has $\Stot=N$.
With \rlb{GS1} in mind, one finds for $M=-N,\ldots,N-1$ that $\Psi_M:=(\Stot^-)^{N-M}\PGS_N$ is nonvanishing and admits the expansion as in \rlb{GS1} with nonnegative coefficients.
Noting that $\beta_{\bsn}>0$ in \rlb{GS1}, we find $\sbkt{\Psi_M,\PGS_M}\ne0$.
But we know that $\Psi_M$ is an eigenstate of $H$ (since $[H,\Stot^-]=0$), and the uniqueness of the ground state within $\calH_M$ shows that $\Psi_M=(\text{const.})\PGS_M$.
Obviously $\Psi_M$ has $\Stot=N$ since $[(\mathbf{S}_\mathrm{tot})^2,\Stot^-]=0$.

It remains to prove the claim about the matrix elements.
Note that the off-diagonal matrix elements $\sbkt{\Phin,H\Phi_{\bsn'}}$ in a bosonic system can  be directly read off from the representation of $H$ in terms of $\ad$ and $a$'s.
This is in a marked contrast with fermionic systems, in which intricate behavior of fermionic sign plays an essential and nontrivial role.
Clearly the hopping term yields a matrix element $-t_{x,y}$ which is non-positive.
To examine the spin-dependent interaction, we recall that 
$(\mathbf{S}_x)^2=(S^+_xS^-_x+S^-_xS^+_x)/2+(S^{(\mrz)})^2$, in which the second term only gives diagonal matrix elements of $H$ in the $\Phin$ basis.
Noting that $S^+_x=\sqrt{2}(\ad_{x,+}a_{x,0}+\ad_{x,0}a_{x,-})$, 
$S^-_x=\sqrt{2}(\ad_{x,0}a_{x,+}+\ad_{x,-}a_{x,0})$, 
an inspection shows that the only contribution to off-diagonal matrix elements comes from $U_2\{\ad_{x,+}\ad_{x,-}(a_{x,0})^2+\text{h.c.}\}$.
This has the desired sign if $U_2\le0$.

To show the connectivity, fix a site $x$.
From any configuration $\bsn$, one first brings all the bosons to $x$.
Then by applying the $U_2$ term, one can change the spin-components to any desired one (within the fixed $M$).
This shows that any configuration in $\calI_M$ is connected (via nonvanishing matrix elements of $H$) to the one configuration where all the particles sit on $x$.

\midskip
\paragraph*{Proof of Theorem 2:}
It is apparent that the $U_2$ terms now produce positive matrix elements $\sbkt{\Phin,H\Phi_{\bsn'}}$, and the Perron-Frobenius theorem does not apply as in the above proof.
Instead we define new basis states by
$\Phin':=(-1)^{\sum_{x}n_{x,0}/2}\,\Phin$, and show that $\sbkt{\Phin',H\Phi'_{\bsn'}}\le0$ for any $\bsn,\bsn'$.
To see this first note that the hopping term again yields the matrix element $-t_{x,y}$ which has the desired sign, since the hopping preserves $\sum_{x}n_{x,0}$.
Next note that the off-diagonal contributions from the $U_2$ terms change the number of spin 0 bosons by two.
Thus the prefactor $(-1)^{\sum_{x}n_{x,0}/2}$ yields an extra $-1$, and the matrix element becomes $-U_2$, which has the desired sign.
Since it is obvious (from the proof of Theorem 1) that the whole set $\calI_M$ are connected via nonvanishing $\sbkt{\Phin',H\Phi'_{\bsn'}}$, we can apply the Perron-Frobenius theorem to conclude that the ground state in $\calH_M$ is unique and is written as \rlb{GS2}.
We note that this argument is the same as that used for $S=1$ spin systems in \cite{spinPF1,spinPF2,spinPF3}, but seems to be more straightforward in the present context of many boson system.

Since $H$, $\Stot^{(\mrz)}$, and $(\mathbf{S}_\mathrm{tot})^2$ are simultaneously diagonalizable, the uniqueness of the ground state (in $\calH_M$) implies that $\PGS_M$ is an eigenstate of $(\mathbf{S}_\mathrm{tot})^2$.
To determine the eigenvalue, consider a toy model  on the same lattice $\La$, with the same boson number $N$, and the Hamiltonian obtained by setting $t_{x,y}=0$ for all $x,y\in\La$, and $V_x=0$ for all $x\in\La\backslash\{o\}$ in $H$ of \rlb{H}, where $o$ is a fixed site in $\La$.
We still have $U_2>0$.
When $V_o$ is negative and large enough, the toy model has ground states in which all the $N$ particles occupy $o$ and are coupled to minimize $(\mathbf{S}_o)^2$.
The bosonic symmetry of the state implies that the minimum possible $\Stot$ in $\calH_M$ is $|M|$ or $|M|+1$ when $N-M$ is even or odd, respectively.
On the other hand, the Perron-Frobenius theorem can be applied also to the toy model to show that the ground state $\tilde{\Psi}^\mathrm{GS}_M$ in $\calH_M$ is unique and admits an expansion like \rlb{GS2} with $\beta_{\bsn}\ge0$.
This means $\sbkt{\tilde{\Psi}^\mathrm{GS}_M,\PGS_M}\ne0$, which implies that $\PGS_M$ has the same $\Stot$ as $\tilde{\Psi}^\mathrm{GS}_M$.

The symmetry $\EGS_M=\EGS_{-M}$ is obvious.
To prove the ordering of $\EGS_M$, take  $M\ge0$ and define $\Psi:=\Stot^{-}\PGS_{M+1}\in\calH_{M}$.
By using $\Stot^+\Stot^-=(\mathbf{S}_\mathrm{tot})^2-(\Stot^{(\mrz)})^2+\Stot^{(\mrz)}$, we find $\norms{\Psi}^2=\sbkt{\PGS_{M+1},\Stot^+\Stot^-\PGS_{M+1}}=2(2M+3)\norms{\PGS_{M+1}}^2$ if $N-M$ is even, 
and $\norms{\Psi}^2=2(M+1)\norms{\PGS_{M+1}}^2$ if $N-M$ is odd.
We thus see $\Psi\ne0$.
Note that $\Psi\in\calH_M$ is an eigenstate of $H$ with the eigenvalue $\EGS_{M+1}$ because $[\Stot^-,H]=0$.
We thus find $\EGS_M\le\EGS_{M+1}$.
When $N-M$ is even, $\Psi$ (like $\PGS_{M+1}$) has $\Stot=M+2$ while $\PGS_M$ has $\Stot=M$.
Thus $\PGS_M$ and $\Psi$ are orthogonal, and the uniqueness of the local ground state in $\calH_M$ implies $\EGS_{M}<\EGS_{M+1}$. 
When $N-M$ is odd, define $\Psi':=\Stot^{+}\PGS_{M}\in\calH_{M+1}$.
By using $\Stot^-\Stot^+=(\mathbf{S}_\mathrm{tot})^2-(\Stot^{(\mrz)})^2-\Stot^{(\mrz)}$ and recalling that $\PGS_{M}$ has $\Stot=M+1$, we find 
$\norms{\Psi'}^2=\sbkt{\PGS_{M},\Stot^-\Stot^+\PGS_{M}}=2(M+1)\norms{\PGS_{M}}^2\ne0$.
Since $\Psi'\in\calH_{M+1}$ is an eigenstate of $H$ with the eigenvalue $\EGS_{M}$, we find  $\EGS_{M+1}\le\EGS_{M}$.
This, with the previous inequality, implies the desired equality $\EGS_M=\EGS_{M+1}$.

\midskip
\paragraph*{Proof of Theorem 3:}
The proof is essentially the same as that of Theorem 1.
We use the fact that $\sbkt{\Phin,H\Phi_{\bsn'}}\le0$ for any $\bsn,\bsn'$, and that the set $\calI_{U,D,Z}$ are connected via nonvanishing $\sbkt{\Phin,H\Phi_{\bsn'}}$.
Then the Perron-Frobenius theorem proves the uniqueness of the local ground state in each $\calH_{U,D,Z}$ and the property \rlb{GS3}.
To show the degeneracy, we again start from the fully polarized ground state $\PGS_{N,0,0}$ and observe that $(\Stot^{+\to-})^D(\Stot^{+\to0})^Z\PGS_{N,0,0}$ is a constant times $\PGS_{N-D-Z,D,Z}$.
Here we have introduced two lowering operators $\Stot^{+\to-}:=\sum_x\ad_{x,-}a_{x,+}$ and  $\Stot^{+\to0}:=\sum_x\ad_{x,0}a_{x,+}$, which commute with $H$ if $U_2=0$.

\midskip
\paragraph*{Some extensions:}
One can replace the interaction part in \rlb{H} by
$\sum_{x\in\La}[
u^{(0)}_xn_x(n_x-1)/2+u^{(2)}_x\{(\mathbf{S}_x)^2/2-n_x\}]$, thus making the interaction site dependent.
The conditions for the theorems then read ``if $u^{(2)}_x\le0$ for any $x$ and $u^{(2)}_x<0$ for some $x$'' for Theorem~1, ``if $u^{(2)}_x\ge0$ for any $x$ and $u^{(2)}_x>0$ for some $x$'' for Theorem~2, and  ``if $u^{(2)}_x=0$ for any $x$'' for Theorem~3.
$u^{(0)}_x$ is arbitrary.

One can take into account non-local and many-body interactions to our theory without any modifications.
The spin-independent interaction $U_0n_x(n_x-1)/2$ can be replaced by any function $U((n_x)_{x\in\La})$ of the number operators, and still Theorems 1, 2, and 3 are valid as they are.
If one replaces the spin-dependent interaction  $U_2(\mathbf{S}_x)^2$ by $\sum_{x,y}J_{x,y}\mathbf{S}_x\cdot\mathbf{S}_y$, Theorem 1 is still valid under the condition $J_{x,y}\le0$ (and $J_{x,y}<0$ for some $x,y$), but Theorem 2 no longer holds under any conditions.
This is because $\mathbf{S}_x\cdot\mathbf{S}_y$ with $x\ne y$ generates an off-diagonal term like $\ad_{x,0}\ad_{y,+}a_{x,+}a_{y,0}$, whose sign is not  changed by the introduction of the prefactor $(-1)^{\sum_{x}n_{x,0}/2}$.

When an external magnetic field in the z-direction is applied to the system, one should add to $H$ the new terms $\sum_x[ -p_x\{n_{x,+}-n_{x,-}\}+q_x\{n_{x,+}+n_{x,-}\}]$, which are the linear and quadratic Zeeman terms, respectively \cite{Zeeman}.
With this modification, which breaks the global SU(2) symmetry, we can only prove  the basic properties \rlb{GS1}, \rlb{GS2}, and \rlb{GS3}, as well as the uniqueness of the ground state in each subspace.

For the system of bosons with spin $S\ge2$ and the same Hamiltonian \rlb{H}, one can prove theorems corresponding to Theorems~1 and 3 in the same manner.
Now the model at $U_2=0$ has a higher SU($2S+1$) symmetry, and the degeneracy becomes larger.
There seems to be no straightforward extension of Theorem~2, suggesting richer phase structures in higher $S$ models \cite{weak_Koahi_Ueda, Ueda_Koashi, Ciobanu}.

\midskip
\paragraph*{Continuous systems:}
Since our theorems cover models with a wide range of parameter values, it is expected that they are valid in the formal continuum limit.
More precisely consider a system of $N$ spin-1 bosons in $\bbR^3$ with the standard Hamiltonian \cite{Ho, Machida}
\eqa
&H=\int d^3\bsr\sum_\sigma\Bigl[
-\frac{\hbar^2}{2m}\Psi_\sigma^\dagger(\bsr)\Delta\Psi_\sigma(\bsr)
+V(\bsr)\,\Psi_\sigma^\dagger(\bsr)\Psi_\sigma(\bsr)
\Bigr]
\nl
&+\int d^3\bsr\,d^3\bsr'
\Bigl[\sum_{\sigma,\sigma'}
U_0(\bsr,\bsr')\Psi_\sigma^\dagger(\bsr)
\Psi_{\sigma'}^\dagger(\bsr')\Psi_\sigma(\bsr)
\Psi_{\sigma'}(\bsr')
\nl
&
+\hspace{-5pt}\sum_{\sigma,\sigma',\tau,\tau'}\hspace{-5pt}
U_2(\bsr,\bsr')\,(\mathbf{S}_{\sigma,\tau}\cdot\mathbf{S}_{\sigma',\tau'})
\Psi_\sigma^\dagger(\bsr)
\Psi_{\sigma'}^\dagger(\bsr')\Psi_\tau(\bsr)
\Psi_{\tau'}(\bsr')
\Bigr]
\ena
where the trap potential $V(\bsr)$ and the spin-independent interaction $U_0(\bsr,\bsr')$ are completely arbitrary.
Then we state (i)~when $U_2(\bsr,\bsr')\le0$ for any $\bsr$ and $U_2(\bsr,\bsr')<0$ for some $\bsr$, all the statements of Theorem~1 are valid as they are; (ii)~when $U_2(\bsr,\bsr')=c_2\,\delta(\bsr-\bsr')$ with $c_2>0$, all the statements of Theorem~2 are valid as they are.

One can easily construct ``proofs'' based on the standard variational argument (see, e.g., \cite{Elliott,EisenbergLieb}) of these statements.
We shall not, however, claim that we have proved them rigorously since mathematically rigorous treatment of continuous systems requires some nontrivial preparations.
In fact the claim (i) may be justified by standard techniques (see, e.g., \cite{RS}) with suitable assumptions on the potentials.
But a rigorous justification of the claim (ii) may require nontrivial effort because of the delta function interaction.

\midskip
\paragraph*{Discussions:}
We have proved basic theorems about the ground states of the standard $S=1$ Bose-Hubbard model.
The result is surprisingly sharp, and we were able to completely characterize the finite-volume ground states for $U_2<0$, $U_2>0$, and $U_2=0$.
A key to understand this sharpness may be the large degeneracy at $U_2=0$,  which reflects the SU(3) symmetry.
Highly degenerate ``SU(3)-ferromagnetic'' ground states at $U_2=0$ are easily lifted either to give ferromagnetic states for $U_2<0$ or a spin-singlet for $U_2>0$ (when the boson number is even).

The robustness of Theorems~1, 2, and 3 and their proofs are also impressive if one recalls that magnetic properties of the Fermi-Hubbard model depend crucially on lattice structures and electron numbers.
Recall, for example, that Lieb's theorem \cite{Hubb_review_Lieb,LiebsTHeorem} (which is a counterpart of Theorem~2) is valid only for the model on a bipartite lattice  at half-filling (and its proof is quite ingenious).
This is a manifestation of the fact that ``exchange interaction'' in bosons are much more tractable than those in fermions.

A remaining challenge is to take these advantages to give a concrete characterization of nontrivial magnetic structures for $U_2>0$ which are intrinsic to spinor bosons.

\bigskip
It is a pleasure to thank 
Takuya Hirano,
Tohru Koma,
Fumihiko Nakano,
Masahiro Takahashi,
Akinori Tanaka,
and 
Yuta Toga
for valuable discussions.



\begin{thebibliography}{10}
\bibitem{eta}
The state with $\Stot=N$ is written in the mean-field form $\Phi=\text{const.}\,P_0(\sum_\sigma \zeta_\sigma b^\dagger_\sigma)^N\Phi_\mathrm{vac}$, where $P_0$ is the projection onto $\calH_0$, and $\zeta_+$, $\zeta_0$, $\zeta_-$ are arbitrary complex constants that satisfy $2\zeta_+\zeta_-=(\zeta_0)^2$ (this is the condition that $(\zeta_+,\zeta_0,\zeta_-)^\mathrm{T}$ is the eigenstate of a spin operator in some direction).
The state with $\Stot=0$ is written as $\Phi=\text{const.}\,\{2b^\dagger_+b^\dagger_--(b^\dagger_0)^2\}^{N/2}\Phi_\mathrm{vac}$, but never in a mean-field form.
\end{thebibliography}
\end{document}